\documentclass[aps,prl,twocolumn,superscriptaddress,showpacs,preprintnumbers,amsmath,amssymb]{revtex4}
\usepackage{graphicx}
\usepackage{amsmath}
\usepackage{amssymb}

\begin{document}
\title{Light transport in cold atoms and thermal decoherence}
\date{\today}

\author{G. Labeyrie}
\email[]{Guillaume.Labeyrie@inln.cnrs.fr}
\affiliation{Institut Non Lin\'{e}aire de Nice,
UMR 6618 CNRS, 1361 route des Lucioles, F-06560 Valbonne, France}
\author{D. Delande}
\affiliation{Laboratoire Kastler Brossel, Universit\'{e} Pierre et Marie Curie,
4 Place Jussieu, F-75005 Paris, France}
\author{R. Kaiser}
\affiliation{Institut Non Lin\'{e}aire de Nice,
UMR 6618 CNRS, 1361 route des Lucioles, F-06560 Valbonne, France}
\author{C. Miniatura}
\affiliation{Institut Non Lin\'{e}aire de Nice,
UMR 6618 CNRS, 1361 route des Lucioles, F-06560 Valbonne, France}
\affiliation{Department of Physics, National University of Singapore, Singapore 117542}

\begin{abstract}
By using the coherent backscattering interference effect, we investigate experimentally and theoretically how coherent transport of
light inside a cold atomic vapour is affected by the residual motion of atomic scatterers.
As the temperature of the atomic cloud increases, the interference contrast dramatically decreases
emphazising the role of motion-induced decoherence for resonant scatterers even in the
sub-Doppler regime of temperature. We derive analytical expressions for the corresponding coherence time.
\end{abstract}

\maketitle

Originally motivated by astrophysical purposes, wave transport in opaque media was first analyzed
through a detailed balance of energy transfers within the scattering medium, leading to radiative
transfer equations~\cite{Chandra}. One main physical ingredient of this theory is a random-phase assumption,
thereby discarding all possible interference effects. In its simplest form, this theory predicts a
diffusive transport with spatial diffusion constant $D=\ell^2/3\tau$, where $\ell$ is the scattering mean free path
and $\tau$ the transport time. However, it is now clear that interference can survive disorder average
in the \textit{elastic} regime and dramatically alter wave transport, leading for example to the
physics of weak and strong localization~\cite{LesHouches}. There has thus been much effort
in solid state physics to study, master and circumvent any possible phase-breaking mechanism to fully access
the regime of \textit{coherent} transport~\cite{SpinFlip}. Coherence loss phenomena can be characterized by
a phase-breaking time $\tau_\Phi$ and a corresponding coherence length $L_{\Phi}=\sqrt{D\tau_\Phi}$~\cite{LesHouches}
beyond which interference effects
are essentially washed out. The coherent transport regime (also known as the \textit{mesoscopic} regime)
is then reached when the medium size $L$ is smaller than $L_\Phi$. For instance, in metals or
semi-conductors, it prevails up to $L_{\Phi}
\approx 1-10 \mu$m even at very low temperatures and very pure samples,
thus requiring the use of devices with micrometer size~\cite{LesHouches}.

During the last two decades, the field got a renewed attention
in systems using electromagnetic waves~\cite{Poan} and, among the large variety of scattering media employed so far,
cold atomic vapors have recently emerged. An important feature of these media is the sharp resonance
of the scattering cross-section (of order $\lambda^2$ on resonance, $\lambda$ being the wavelength),
allowing for a continuous tuning of the scattering mean free path $\ell$ via the light frequency.
Phase-breaking mechanisms in cold atomic vapors can be efficiently probed by using the coherent
backscattering effect (CBS), a paradigmatic two-wave interference effect in multiple scattering~\cite{CBS}.
In the limit of low light intensity, an important mechanism has been identified~\cite{ThmeetExp}.
It is rooted in the Zeeman degeneracies of the
atomic internal structure but could be healed by applying a magnetic field to the sample~\cite{F2champB}.
Another important phase-breaking mechanism is the residual thermal motion
of atomic scatterers inside the cold vapor and is the main topic of this paper.
Indeed, because of the Doppler and recoil effects, light frequency is changed upon scattering and
transport is \textit{no more} elastic. In previous papers~\cite{radtrap1,radtrap2}, we have shown how the incoherent
transport is affected by such effects. For typical alkali magneto-optical traps (MOT),
Doppler-induced frequency shift is the dominant effect.
Not surprisingly, an important
parameter is the ratio of the typical Doppler shift $kv$
($v$ being the 1D rms velocity, and $k=2\pi/\lambda$ the
incoming light wavevector) to the width $\Gamma$ of the atomic resonance.
When $kv \gg \Gamma$, a single scattering
event is enough to bring the photon completely out of resonance thereby altering transport significantly~\cite{David}.
When $kv \ll \Gamma$, a single scattering event only slightly modifies the photon frequency and leads,
as a cumulative effect, to diffusion in frequency space~\cite{radtrap2}.

In this Letter, we report the first unambiguous experimental evidence of a phase-breaking
mechanism in light transport induced
by the residual atomic motion~\cite{Fink}. We first describe the experimental procedure and our
main result. We then give an analysis of the physical ingredients at the heart of the light coherence reduction.
We finally compare our data to the Monte-Carlo simulations described in~\cite{BouleG},
appropriately modified to include the effect of the atomic velocity distribution~\cite{Kupriyanov}.

\begin{figure}
\begin{center}
\includegraphics[width=7cm]{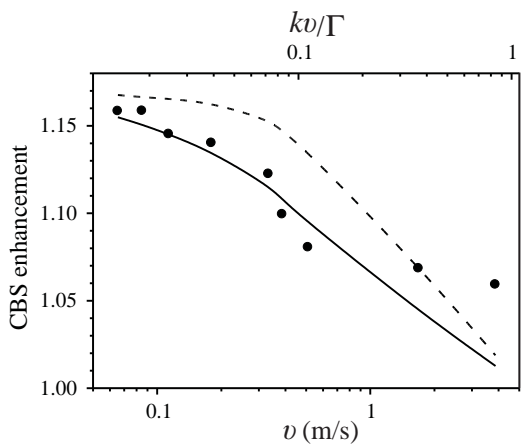}
\end{center}
\caption{\label{fig1}
CBS enhancement factor (full circles) as a function of the 1D typical atomic velocity $v$ and its comparison
to a Monte-Carlo calculation assuming either a Gaussian (dashed line) or a Lorentzian (solid line)
atomic velocity distribution.
The CBS peak is recorded in the helicity non-preserving channel, with resonant laser light ($\delta=0$).
The optical thickness of the cloud is set constant at $b \simeq 13$. For Rubidium,
the velocity scale $\Gamma/k$ is 4.6 m/s.}
\end{figure}

\begin{figure}
\begin{center}
\includegraphics[width=6cm]{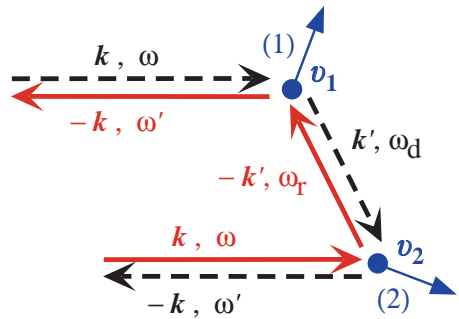}
\end{center}
\caption{\label{fig2}
(color online) CBS dephasing induced by the atomic motion, in the simplest case of two atoms moving
with different velocities ${\mathbf v}_1$ and ${\mathbf v}_2$.
The CBS effect builds on the two-wave interference between amplitudes associated to reversed
multiple scattering paths (arrows). The Doppler effect induces frequency redistribution of the scattered light.
Although the outgoing
frequencies are identical, the intermediate frequencies differ, yielding a phase coherence loss and
thus a reduction of the CBS contrast.}
\end{figure}

\begin{figure}
\begin{center}
\includegraphics[width=7cm]{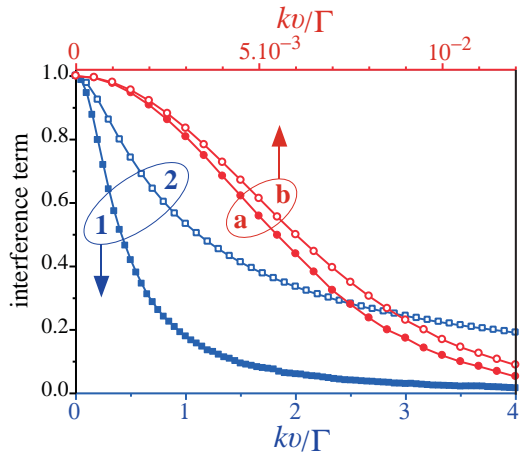}
\end{center}
\caption{\label{fig3}
(color online) Numerical calculation of the CBS interference decay for multiple scattering order
$N=2$ (squares) and $N=40$ (circles).
The upper velocity range refers to curves {\bf a} and {\bf b}, while the lower one refers
to curves {\bf 1} and {\bf 2}. Note the very different velocity ranges. We
compare two situations: resonant excitation $\delta=0$ ({\bf 1} and {\bf a}) and detuned excitation
$\delta \gg kv$ ($\delta=50 \Gamma$ for {\bf 2} and $\delta=5 \Gamma$ for {\bf b}).
Although the CBS contrast is improved, a detuned excitation is not sufficient to circumvent
the motion-induced decoherence.}
\end{figure}

Our CBS experimental setup has been described in
details elsewhere~\cite{BouleG}, the main difference here being the heating procedure to increase $v$.
It is achieved by switching off the magnetic gradient of the MOT and then exposing the cloud
to a 200$\mu$s long, slightly red-detuned, optical molasses. An increasing amount
of heating is obtained by tuning the molasse frequency closer to resonance. The resulting
3D velocity distribution is measured by releasing atoms from the trap and imaging their ballistic
expansion. With some word of caution (one side-effect of this heating technique is to change the atomic
velocity distribution which can differ from a Gaussian one), the temperature of the cloud is $k_BT = mv^2$.
The number of atoms in the cloud can be adjusted independently of $T$ so as to maintain a fixed optical
thickness $b=L/\ell$. In Fig.~\ref{fig1} we observe a fast decrease of the CBS enhancement factor
as $v$ is increased, the value obtained as $v \to 0$ being set by the Rubidium Zeeman degeneracy.
In particular, the decay is faster than
expected from the naive criterion $k v \approx \Gamma$.
The measurements are performed in the helicity non-preserving channel
with a resonant laser (detuning $\delta=\omega-\omega_0=0$, where $\omega$ and $\omega_0$ are
the laser and atomic angular frequencies).
The measured angular width of
the CBS peaks $\Delta\theta \simeq 1/k\ell$ ($k \ell \simeq 1000$
typically) varied by less than $10\%$ in the
course of the experiment, which proves that the reported decrease is not driven
by the angular resolution of the apparatus.
Before commenting Fig.~\ref{fig1}, let us discuss the physical ingredients of
motion-induced CBS reduction for the simple double scattering configuration shown in Fig.~\ref{fig2}.
We assume here that the atomic motion is classical~\cite{mueller05}. Atoms, distributed in space with number
density $\rho \ll k^3$, are illuminated by a plane wave of angular frequency $\omega$ and
wave vector ${\mathbf k}$ and build up a dilute effective medium characterized by its
complex index of refraction $n \approx 1+\rho \alpha /2$, convolved by the atomic velocity distribution.
The atomic complex polarizability is $\alpha = -3\pi\Gamma k^{-3}/(\delta+\text{i}\Gamma/2)$.
The CBS signal can be understood as an interference between a
``direct" scattering path (solid arrows) and its ``reverse" counterpart (dashed arrows).
To each path is associated a complex amplitude both incorporating scattering and propagation in the effective medium.
For scatterers at rest, with no Zeeman degeneracies, these amplitudes are exactly balanced at backscattering
(outgoing wave vector $-{\mathbf k}$) yielding full CBS contrast.
For Zeeman degenerate atoms, like Rubidium, we recover the results of~\cite{ThmeetExp}.
Let us now consider the impact of the atomic motion. Because of
the Doppler effect, the light frequency in
the laboratory reference frame is modified. For the direct path, it is
$\omega_d=\omega+({\mathbf k}^{\prime}-{\mathbf k}).{\mathbf v}_1$ where $\mathbf k^{\prime}$ is the intermediate
wavevector.
After scattering by atom 2, the outgoing frequency is
$\omega^{\prime}=\omega+({\mathbf k}^{\prime}-{\mathbf k}).{\mathbf v}_1-({\mathbf k}+{\mathbf k}^{\prime}).{\mathbf v}_2$.
For the reverse path, the frequency between the two atoms is
$\omega_r = \omega-({\mathbf k}^{\prime}+{\mathbf k}).{\mathbf v}_2 \neq \omega_d$, while
the final frequency is again $\omega^{\prime}$. Thus, although the outgoing frequencies are identical and do interfere,
the intermediate frequencies differ by an amount $\simeq kv$, resulting in different complex amplitudes
and reducing on average the CBS interference contrast. Similarly,
the frequencies seen by the atoms in their rest frames are different along the direct
and reverse paths, leading to different scattering phaseshifts. More specifically,
for small velocities $kv \ll \Gamma$ one can estimate the typical phase difference
$\Delta\Phi$ induced by the atomic motion between
the reversed paths:
\begin{equation}
\Delta\Phi=\Delta\Phi_s + \Delta\Phi_p \simeq [\partial_\omega\Phi_s+
k\ell\,\partial_\omega n]\, kv \approx kv/\Gamma
\label{phase}
\end{equation}
The first term corresponds to the scattering phaseshift, while the second describes the phaseshift
associated to propagation over a distance $\ell$ in the effective medium.
Although approximate, eq.~(\ref{phase}) provides useful insights about the impact of $\delta$ at finite $v$ (see below).
For larger scattering orders ($N > 2$), the elementary scattering and propagation processes have to be chained
and randomly distributed phase differences will accumulate, decreasing further the CBS contrast, see Fig.\ref{fig3}.
In order to accurately compute the shape and height of the CBS peak, one must average over all atomic degrees of freedom,
both external (position and velocity) and internal (magnetic quantum number),
but one must also incorporate the experimental geometry such as the shape, size, optical thickness of the cloud as
well as the waist and linewidth of the CBS probe laser beam. A Monte-Carlo approach as described in~\cite{BouleG}
is then necessary, where we now include the effect of the atomic velocity distribution~\cite{radtrap1}.
As mentioned earlier, some of the measured velocity distributions exhibited clear deviations from a Gaussian with an
excess of fast atoms.
Such atoms are responsible for large Doppler effects and are thus likely
to kill CBS more efficiently. We have used in our simulations either a purely Gaussian velocity distribution
or a Lorentz-like distribution (where we
define $v=v_0/(\sqrt{8ln2})$, $v_0$ being the FWHM-velocity).
The results, displayed in Fig.~\ref{fig1}, show a good qualitative agreement,
with a fast decay of the enhancement factor with increasing $v$.
However, the Gaussian distribution show deviations at low $v$ while the Lorentz distribution
gives a better agreement. This pinpoints the important role
played by a small fraction of atoms with large velocities.
A quantitative agreement would require a better knowledge and control
of the initial atomic velocity distribution.
As can be seen, the CBS contrast decreases faster  than naively expected.
This is because higher scattering orders contribute significantly to the CBS effect.
Indeed, at large optical thickness ($b > 10$), they contribute altogether
to $\approx 30 \%$ of the CBS peak height. As increasing scattering orders are more and more sensitive
to motion-induced phase-breaking, this explains the reported observations.

Since the CBS scrambling originates from the frequency-dependent response of the
sample (which is maximum around resonance), one
could expect to recover a full interference contrast by detuning the laser at $|\delta | \gg kv$.
This is an important issue for the prospective CBS observation using a room-temperature vapor.
Fig.~\ref{fig3} displays Monte-Carlo simulations of the CBS contrast as a function of the rms
velocity $v$ at $\delta=0$ and $\delta \gg kv$. Calculations are made for
$N=2$ (squares)
and $N=40$ (circles) in a semi-infinite medium.
As can be seen (curves {\bf a} and {\bf b}, $N=40$), at low velocities $kv \ll \Gamma$
the interference contrast is nearly frequency independent.
This behavior is due to a compensation between the variations of the scattering and propagation terms.
Indeed, the phase difference in eq.~(\ref{phase}) is essentially proportional to the transport time
$\tau$ (bracketed term), which was shown to be frequency-independent near
the resonance and equal to $\Gamma^{-1}$~\cite{radtrap1}. As mentioned earlier and expressed in
eq.~(\ref{attenuation}), the decay rate is faster for
higher scattering orders.
At larger velocity (curves {\bf 1} and {\bf 2}, $N=2$), Fig.~\ref{fig3} shows an increased frequency dependence
of the interference contrast, which we interpret as an unperfect balance between the scattering and
propagation phaseshifts in eq.~(\ref{phase}). The interference contrast is partially
restored in the $|\delta| \gg kv$ limit, where the motion-induced decoherence is now essentially driven
by the propagation term in eq.~(\ref{phase}) of
order $kv/\Gamma$. The interference thus still decreases towards zero with increasing $v$.

Beyond the specific case of CBS, one can quantitatively characterize the motion-induced phase breaking by
defining the phase-breaking time $\tau_\Phi$ and its associated coherence length $L_{\Phi}$.
We discuss here the simplest low-velocity case, $kv \ll \Gamma$.
At each scattering event, the photon detuning is slightly modified and this cumulative process turns into a
random walk at sufficiently large scattering orders $N \gg 1$. Its dispersion is
$\approx kv\,\sqrt{N}$ which we also assume to be $\ll \Gamma$.
If the atomic velocities are Gaussian distributed
and independent, the averaging over the degrees of freedom can be performed analytically, and the resulting
interference contrast decreases like:
\begin{equation}
c_N(v) = \exp{\left[-\frac{N^3}{12} \left(\frac{kv}{\Gamma}\right)^2\right]}
\label{attenuation}
\end{equation}
Note that this quantity is independent of the initial detuning
$\delta$ as discussed before. This expression is also valid for larger velocities,
in the off-resonant limit $\delta \gg \sqrt{N}\,kv$ where the
convolution by the velocity distribution can be neglected.
The decay of the interference contrast described by eq.~(\ref{attenuation}) is not exponential
with $N,$ in contrast to other sources of decoherence like the atomic internal structure~\cite{ThmeetExp}.
Nevertheless,
one can still define a critical scattering order $N_\Phi = 3 (3kv/2\Gamma)^{-2/3}$. This means that the
motion-induced phase-breaking mechanism can be characterized by:
\begin{subequations}
\begin{eqnarray}
\label{Tphi}
\tau_\Phi &=& 3 \left(\frac{3}{2}\frac{kv}{\Gamma}\right)^{-2/3} \; \Gamma^{-1}\\
L_{\Phi} &=& \left(\frac{3}{2}\frac{kv}{\Gamma}\right)^{-1/3}\; \ell
\label{Lphi}
\end{eqnarray}
\end{subequations}
Therefore, the mesoscopic condition $L_{\Phi} > L$ reads:
\begin{equation}
\frac{kv}{\Gamma} < \frac{1}{b^3}
\label{Meso}
\end{equation}
For our lowest temperature of $\approx 40\ \mu$K ($v=6.5$ cm/s),
eq.~(\ref{Lphi}) yields $L_{\Phi} \simeq 4\ell$ and
condition~(\ref{Meso}) is fulfilled for optical thicknesses $b\leq
4$.

At this point, an interesting comparison with electronic transport in metals can be made.
There, the temperature dependence of $\tau_{\Phi}$ is governed by several different mechanisms: electron-electron
scattering ($\propto T^{-2/3}$), electron-phonon scattering ($\propto T^{-3}$) and spin-flip scattering
(weakly $T$-dependent) which dominates at low $T$~\cite{SpinFlip,Altshuler}.
In our case, $\tau_{\Phi}$ is set by the internal structure~\cite{ThmeetExp} (independent of $T$) and by the
motion-induced decoherence, which, according to eq.~(\ref{Tphi}),
yields $\tau_{\Phi} \propto T^{-1/3}$. At $T=40\ \mu$K, $\tau_{\Phi}$ is mostly determined by the internal
structure (see Fig.~\ref{fig1}).

A simple dynamical picture of the motion-induced coherence loss has been given by Golubentsev~\cite{Golub}.
In this picture, phase-coherence is lost when the length of a multiple scattering path varies
(due to the motion of the scatterers) by $\lambda$ during the time taken by the
light to follow this path. Golubentsev found expressions equivalent to ours, with
however a huge difference: for the non-resonant case he considered, the transport time is
$\tau = \ell/c < 1$ ps while for our narrow resonance this time is $\Gamma^{-1} \approx 27$ ns,
more than 4 orders of magnitude higher. It is only because photons are ``slowed down'' inside the resonant medium
that we are able to observe the destruction of phase coherence.

In conclusion, we reported the first observation of a motion-induced phase breaking of the CBS effect.
The observed CBS reduction is due to the fast-varying frequency
response of the medium in the vicinity of a resonance, which enhances the impact of frequency redistribution caused by
Doppler effect.
The rapid decrease of the interference contrast indicates
that the phase coherence length $L_{\Phi}$ associated with the residual motion of atoms
in our coldest sample is not very large compared to the scattering mean free path, and smaller than the
actual size of the cloud. According to the mesoscopic criterion derived in this paper,
extremely low temperatures, in the nK range, are probably required to
fully access interference effects in optically-thick cold atomic gases. In this regime however,
the recoil effect is expected to become
the main source of decoherence.

We thank David Wilkowski, Beno\^{\i}t Gr{\'e}maud and Cord M\"uller for useful discussions.
We acknowledge the financial support of CNRS and of the PACA Region. Laboratoire Kastler Brossel is laboratoire
de l'Universit{\'e} Pierre et Marie
Curie et de l'Ecole Normale Sup{\'e}rieure, UMR 8552 du CNRS. CPU time
on various computers has been provided by IDRIS.

\end{document}